\documentstyle[12pt,psfig]{article}

\newcommand{\ns}{E866/NuSea }
\newcommand{\ud}{$\bar{d}-\bar{u}$ }
\newcommand{\udq}{$\bar{d}/\bar{u}$ }
\newcommand{\beq}{\begin{equation}}
\newcommand{\eeq}{\end{equation}}
\newcommand{\bea}{\begin{eqnarray}}
\newcommand{\eea}{\end{eqnarray}}

\begin{document}

\title{{\bf The  $\bar{d}$ $\bar{u}$  asymmetry of 
the proton in a pion cloud model approach}}
\author{{ J. Magnin}\thanks{Present address: Depto. de F\'{\i}sica, 
Universidad de los Andes,
AA 4976, Santaf\'e de Bogot\'a, Colombia. e-mail: 
jmagnin@uniandes.edu.co} 
\ {\small and} 
{H.~R. Christiansen}\thanks{e-mail: hugo@cbpf.br} \\  \\
{\normalsize {\it Centro Brasileiro de Pesquisas F\'{\i}sicas}, 
{\small CBPF}} \\ 
{\normalsize {\it Rua Dr. Xavier Sigaud 150, 22290-180, 
Rio de Janeiro, Brazil}}}

\date{{\tt {To appear in Physical Review D}}}
\maketitle

\begin{abstract}
We study the  $\bar{d}$ $\bar{u}$ asymmetry of the proton in a model
approach recently developed, in which hadronic fluctuations of 
the nucleon are generated through gluon splitting and recombination 
mechanisms. \- Within this framework, it is shown that 
both $\bar{d}/\bar{u}$ and \ud distributions 
in the proton can be consistently described by including only nucleon 
fluctuations to $\left|\pi N\right>$ and $\left|\pi \Delta\right>$ 
bound states. Predictions of the model closely agree 
with the recent experimental data of the \ns Collaboration.
\end{abstract} 

In 1991, the New Muon Collaboration (NMC)~\cite{nmc} presented a determination
of the non-singlet structure function $F_2^p - F_2^n$ at $Q^2 = 4$ GeV$^2$
over the range $0.004<x<0.08$. From this measurement, the Gottfried Sum
Rule (GSR)~\cite{gsr} was estimated and a significantly lower value than the
$1/3$ predicted by the quark-parton model was found.
Although it could be due to an abnormal behavior of the valence quark 
distributions in the unmeasured region~\cite{stirling},
the above result was attributed to a \ud asymmetry in the light nucleon
sea (see e.g. Ref.~\cite{preparata}). Later on, the NA51 Collaboration~\cite{na51} 
determined the value $\bar{d}/\bar{u}=1.961\pm0.252$ at $x=0.18$ in 
Drell-Yan dimuon production, giving strong experimental
support to the \ud asymmetry explanation.

Most recently, the \ns Collaboration measured the ratio \udq in the nucleon
over the range $0.02<x<0.345$ in Drell-Yan dimuon production from
$p-p$ and $p-D$ interactions~\cite{nusea}. From this measurement,
the \ns Collaboration extracted the $x$ dependence of \ud and 
estimated a value for its integral, $\int_0^1{dx[\bar{d}-\bar{u}]}=0.100\pm0.018$,
indicating a strong GSR violation. 
Notably, this value is only about two thirds
of the NMC estimate for the same integral, $0.147\pm0.026$. 
We will address this issue later on.

In conclusion, although a $\bar{d}$ $\bar{u}$ asymmetry in the nucleon sea 
has been 
firmly established from experiments, its origin and precise features are yet unclear.
Several ideas have been put forward to try to explain the GSR violation
and the \ud asymmetry in nucleons. Among them the Pauli exclusion principle, 
which would inhibit the development of up (down) quarks and anti-quarks 
in the proton (neutron) sea, a pioneer idea by Field and 
Feynman~\cite{feynman}; fluctuations of valence quarks into quarks 
plus massless pions~\cite{eichten}, an effect which is calculable 
in Chiral Field Theory; and earlier versions of the Pion Cloud
Model (PCM)~\cite{varios}. However, none of these attempts gave a satisfactory
description of the experimental status of {\it both} \ud and \udq distributions.
It should be noted that previous versions of the PCM used rather
hard pion distributions inside the nucleon resulting in large 
$\bar{u}$ and $\bar{d}$ distributions beyond $x\sim 0.3$.
These large pion contributions can not be easily compensated to 
conveniently describe the fast fall-off of \ud in the whole $x$ range. 
Albeit large for small $x$, the \ud distribution seems to be negligible 
beyond $x\sim 0.3$. 
In addition, the \udq ratio predicted by these models exhibits a dramatic 
growing behavior not seen in the experimental data. 
For reviews, see Ref.\cite{reviews}
\vskip .5cm

In this work we will show that a recently proposed version of the
Pion Cloud Model (PCM)~\cite{ch-m} allows a remarkable prediction of 
the nucleon's $\bar{d}$ $\bar{u}$ asymmetry in accordance with the 
recent results of the \ns Collaboration. 
Our approach is based on both perturbative and effective degrees of 
freedom, and it relies on a recombination model description of the
hadronic fluctuations of the nucleon. 

Let us briefly recall the model introduced in  Ref.~\cite{ch-m}. 
We start by considering a simple picture of the ground state of the 
proton in the infinite momentum frame as formed by three valence quark 
clusters or {\em valons}~\cite{hwa}. The valon distributions in 
the proton are given by 
\begin{equation}
v(x) = \frac{105}{16} \sqrt{x} \left( 1 - x\right)^2 \;,
\label{eq1}
\end{equation}
where, for simplicity, we do not distinguish between $u$ and $d$ 
valons. 

The higher order contributions to the proton structure 
are identified with meson-baryon bound states in 
an expansion of the nucleon wave-function in terms of hadronic Fock states. 
Such hadronic fluctuations are built up by allowing that a valon emits a 
gluon which, before interacting with the remaining valons, decays 
perturbatively into a $q\bar{q}$ pair. This quark anti-quark pair 
subsequently recombines with the valons so as to form a meson-baryon 
bound state.

The probability distributions of the initial perturbative 
$q\bar{q}$ pair can be calculated by means of the 
Altarelli-Parisi~\cite{alta-par} splitting functions:
\beq
P_{gq} (z) = \frac{4}{3} \frac{1+(1-z)^2}{z},\ \ \ \ 
P_{qg} (z) = \frac{1}{2} \left( z^2 + (1-z)^2 \right).
\label{eq2}
\eeq
Accordingly, the joint probability density of obtaining a quark 
or anti-quark coming from  subsequent decays $v \rightarrow v + g$ 
and $g \rightarrow q + \bar{q}$ at some fixed low $Q_v^2$ is 
\begin{equation}
q(x) = \bar{q}(x) = N \frac{\alpha_{st}^2(Q^2_v)}{(2\pi)^2}
\int_x^1 {\frac{dy}{y} P_{qg}\left(\frac{x}{y}\right) 
\int_y^1{\frac{dz}{z} P_{gq}\left(\frac{y}{z}\right) v(z)}} \; .
\label{eq3}
\end{equation}
The value of $Q_v$, as dictated by the valon model of the nucleon, 
is about $Q_v =1$ GeV. For definiteness we take $Q_v =0.8$ GeV as in 
Ref.~\cite{ch-m,hwa}, which is large enough to allow for a perturbative 
evaluation of the $q\bar{q}$ pair production. 
$N$ is a normalization constant whose value depends on the flavor of the 
quark and anti-quarks produced in the $gq\bar{q}$ vertex.

Once $q$ and $\bar{q}$ are created, they may subsequently interact 
with the valons so as to form a most energetically favored 
meson-baryon bound state. The rearrangement of such five-component nucleon 
configuration into a meson-baryon bound state must be evaluated by means 
of effective methods. This is necessary because
the interactions involved in such a process are within the confinement 
region of QCD. Therefore, non-perturbative interactions take place. 
Assuming that the {\em in-proton} meson and baryon 
formation arise from mechanisms similar to those at work in the 
production of real hadrons, we evaluate the in-proton pion probability 
density using a well-known recombination model approach~\cite{das-hwa}. 

Within this scheme, the pion probability density in the 
$\left|\pi B\right>$ fluctuation of the proton is given by
\begin{equation}
P_{\pi B} (x) = \int_0^1 \frac{dy}{y} {\int_0^1\frac{dz}{z} F(y,z) R(x,y,z)},
\label{eq5}
\end{equation}
where $R(x,y,z)$ is the recombination function associated with 
the pion formation, 
\begin{equation}
R(y,z) = \alpha\frac{yz}{x^2} 
\delta \left(1 - \frac{y+z}{x}\right) \; ,
\label{eq7}
\end{equation}
and $F(y,z)$ is the valon-quark distribution function given by
\begin{equation}
F(y,z) = \beta\, yv(y)\, z\bar{q}(z) (1-y-z)^a \; .
\label{eq6}
\end{equation}

The exponent $a$ in eq.~(\ref{eq6}) is fixed by the requirement that the 
pion and the baryon in the $\left|\pi B\right>$ fluctuation have the 
same velocity, thus favoring the formation of the meson-baryon bound 
state. With the above constraint we obtain $a=12.9$ and $a=18$ for the 
$\left|\pi^+ n\right>$ and the $\left|\pi \Delta\right>$ fluctuations 
of the proton respectively.

Note that in the original version of the recombination model 
this exponent was fixed to $1$ \cite{das-hwa}. This is 
basically because in a collision, the only relevant kinematical 
correlation in the model between the initial and final states
is momentum conservation. On the other hand, 
in the present case the recombining quarks are more correlated 
as they are making part of a single object from the outset.
Firstly, meson and baryon must exhaust the momentum of 
the proton~\footnote{We fulfill this requirement by assuming $P_{\pi B}(x)
=P_{B\pi}(1-x)$. See Ref.~\cite{ch-m} for a discussion about this point.}, 
and secondly, they must be correlated in velocity as a bound-state 
is expected to be formed.

The overall normalization $N\beta\alpha$ of the probability 
density $P_{\pi B}$ must be fixed by comparison with experimental data.

The non-perturbative $\bar{u}$ and $\bar{d}$ distributions 
can now be computed by means of the two-level convolution formulas  
\bea
\bar{d}^{NP}(x,Q_v^2) & = & \int^1_x {\frac{dy}{y} 
\left[P_{\pi N}(y) + \frac{1}{6} P_{\pi \Delta}(y) \right] 
\bar{d}_{\pi}(\frac{x}{y},Q_v^2)} \label{eq8a} \\
\bar{u}^{NP}(x,Q_v^2) & = & \int^1_x {\frac{dy}{y} \frac{1}{2} 
P_{\pi \Delta}(y)\ \bar{u}_{\pi}(\frac{x}{y},Q_v^2)},
\label{eq8}
\eea
where the sources $\bar{d}_{\pi}(x,Q_v^2)$ and 
$\bar{u}_{\pi}(x,Q_v^2)$ are the 
valence quark probability densities in the pion at the low $Q_v^2$ 
scale. In eq.~(\ref{eq8a}), we have summed the contributions 
of the $\left|\pi^+ n\right>$ and $\left|\pi^+ \Delta^0\right>$ 
fluctuations to obtain the total non-perturbative $\bar{d}$ 
distribution. For the non-perturbative $\bar{u}$ distribution of 
eq.~(\ref{eq8}), the only contribution originates from the 
$\left|\pi^- \Delta^{++}\right>$ fluctuation. 
Contributions arising from fluctuations containing neutral mesons
as the $\pi^0$, are strongly supressed in this model and will not be 
considered\footnote{This supression is associated with their flavor structure 
(in terms of their parton components). In particular for 
$\pi^0 \simeq (d\bar{d} - u\bar{u})$ fluctuations. 
In our model the first component arises from the recombination of a $\bar d$ quark 
and a $d$ valon, and the second  comes from the recombination of a $\bar u$ quark 
and a $u$ valon. 
Due to their unflavored structure, neutral in-nucleon quark-antiquark objects would 
tend to be annihilated much more rapidly than charged, flavored ones like
$u\bar d$ or $d\bar u$.
In other words, unflavored $q\bar q$ recombination would be highly
inhibited and neutral pion fluctuations would be ephemeral. Thus, 
$\pi^0$ configurations would be unlikely to occur in comparison with  
$\pi^\pm$ fluctuations, although they belong to the same mass multiplet.}.
We also neglect higher order Fock components.

The factors $\frac{1}{6}$ and $\frac{1}{2}$ in front of 
$P_{\pi \Delta}$ in eqs.~(\ref{eq8a}) and (\ref{eq8}) 
are the (squared) Clebsh-Gordan (CG) coefficients needed 
to account for the $\frac{1}{2}$ isospin constraint on the fluctuation. 
The CG coefficient corresponding to the $\left|\pi^+ n\right>$ fluctuation 
is hidden in the global normalization of the state.

We will now compare our results with the experimental data.
As the \ns Collaboration measures the ratio \udq at $Q=7.35$ GeV, 
we first compute this quantity by means of
\beq
\frac{\bar{d}(x,Q^2)}{\bar{u}(x,Q^2)} = 
\frac{\bar{d}^{NP}(x,Q^2) + \bar{q}^P(x,Q^2)}
{\bar{u}^{NP}(x,Q^2) + \bar{q}^P(x,Q^2)}\ .
\label{eq9}
\eeq
Here $\bar{d}^{NP}(x,Q^2)$ and $\bar{u}^{NP}(x,Q^2)$ are given by
eqs.(\ref{eq8a}) and (\ref{eq8}) and
$\bar{q}^P(x,Q^2)$ represents the perturbative part of the up 
and down sea of the proton, which we assume to be equal. 
This assumption is exact up to at least $1\%$~\cite{pert}. 

Regarding the difference $\bar{d}-\bar{u}$, instead of computing it 
directly by subtracting
eqs.~(\ref{eq8a}) and (\ref{eq8}), we will extract it from the \udq ratio 
as done in Ref. \cite{nusea}. In its paper, the \ns Collaboration employed the 
following identity to obtain the difference:
\beq
\bar{d}(x) - \bar{u}(x) = \frac{\bar{d}(x)/\bar{u}(x) - 1}
{\bar{d}(x)/\bar{u}(x) + 1} \ [\bar{u}(x) + \bar{d}(x)]\ .
\label{eq9a}
\eeq
While the ratio $\bar{d}(x)/\bar{u}(x)$ is a direct measurement of E866
the sum $\bar{u}(x) + \bar{d}(x)$ appearing in eq.~(\ref{eq9a}) is taken 
from the CTEQ4M parametrization~\cite{cteq}. 

\begin{figure}[b] 
\psfig{figure=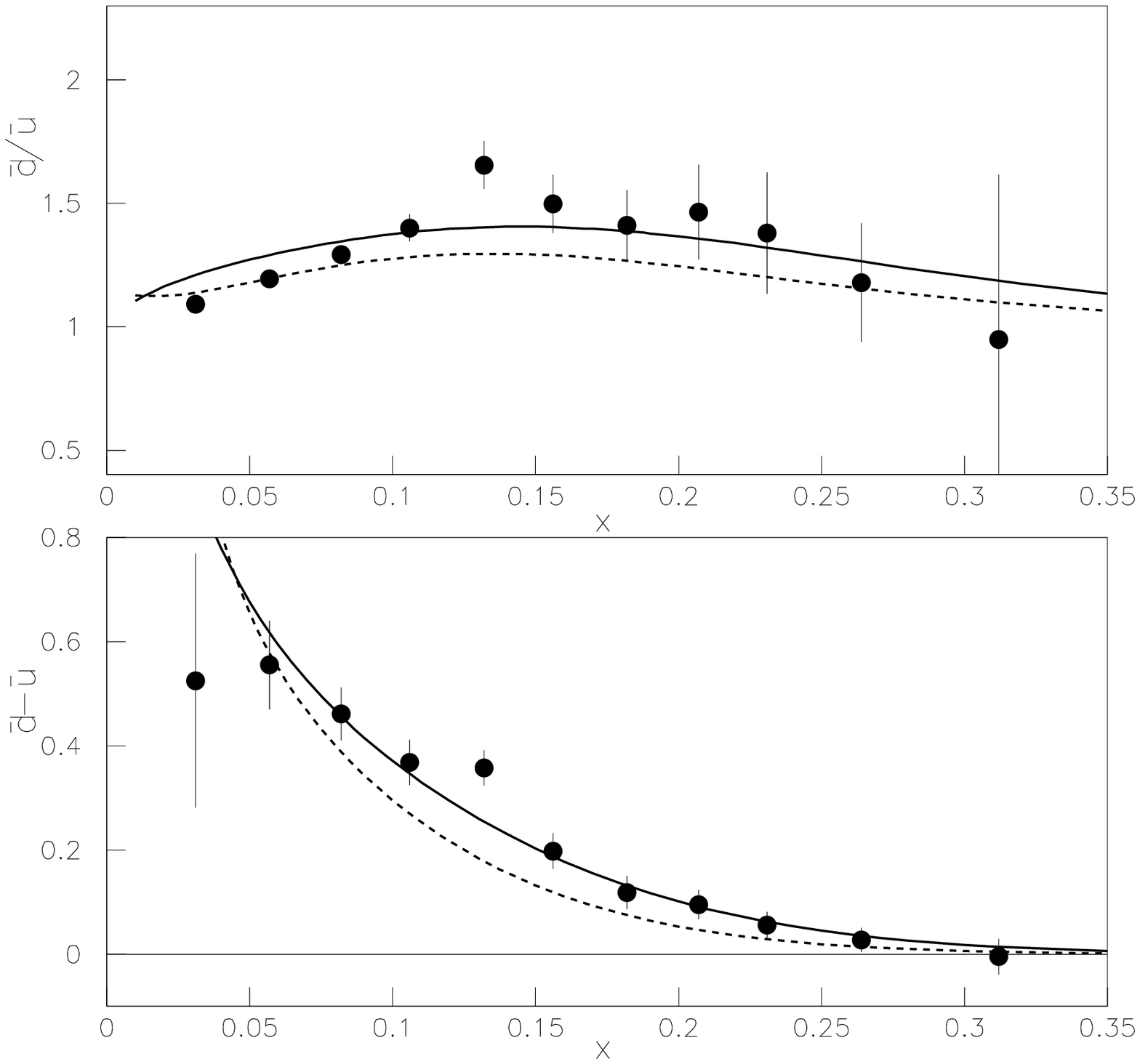,height=6.0in} 
\caption {Predictions of the model compared with experimental data 
from Ref.~\cite{nusea}. \udq ratio (upper) and \ud asymmetry (lower) 
at $Q=7.35$ GeV. Curves are calculated with unevolved 
$\bar{u}^{NP}$ and $\bar{d}^{NP}$ distributions (full line) and 
with pseudo-evolved non-perturbative distributions (dashed line)}
\label{fig1}
\end{figure}

In Fig.~\ref{fig1}, our predictions of \udq and \ud 
are compared with the experimental data from Ref.~\cite{nusea}. 
The curves were obtained using the pion valence 
distributions of Ref.~\cite{grv-pion} in eqs.~(\ref{eq8a}) and 
(\ref{eq8}) and the proton sea quark distributions of Ref.~\cite{grv-p} 
in eq.~(\ref{eq9}). 

Note that a rigorous comparison of our prediction with
the experimental data would require that the non-perturbative 
$\bar{u}$ and $\bar{d}$ distributions be evolved up to $Q = 7.35$ 
GeV. Instead of performing a full QCD evolution program, we 
{\em pseudo-evolve} the $\bar{u}^{NP}$ and $\bar{d}^{NP}$ distributions 
by multiplying them by the ratio $q(x,Q^2=7.35^2$ GeV$^2)/q(x,Q_v^2)$. 
The function $q$ represents the corresponding valence quark distribution 
in the proton at the \ns and the valon scales respectively.
This simple procedure is satisfactory enough to give us a feeling of the
effect of the evolution of the non-perturbative distributions on \udq and
\ud\footnote{A similar strategy has been adopted in Ref.\cite{bro-ma}}. 

\begin{figure}[b] 
\psfig{figure=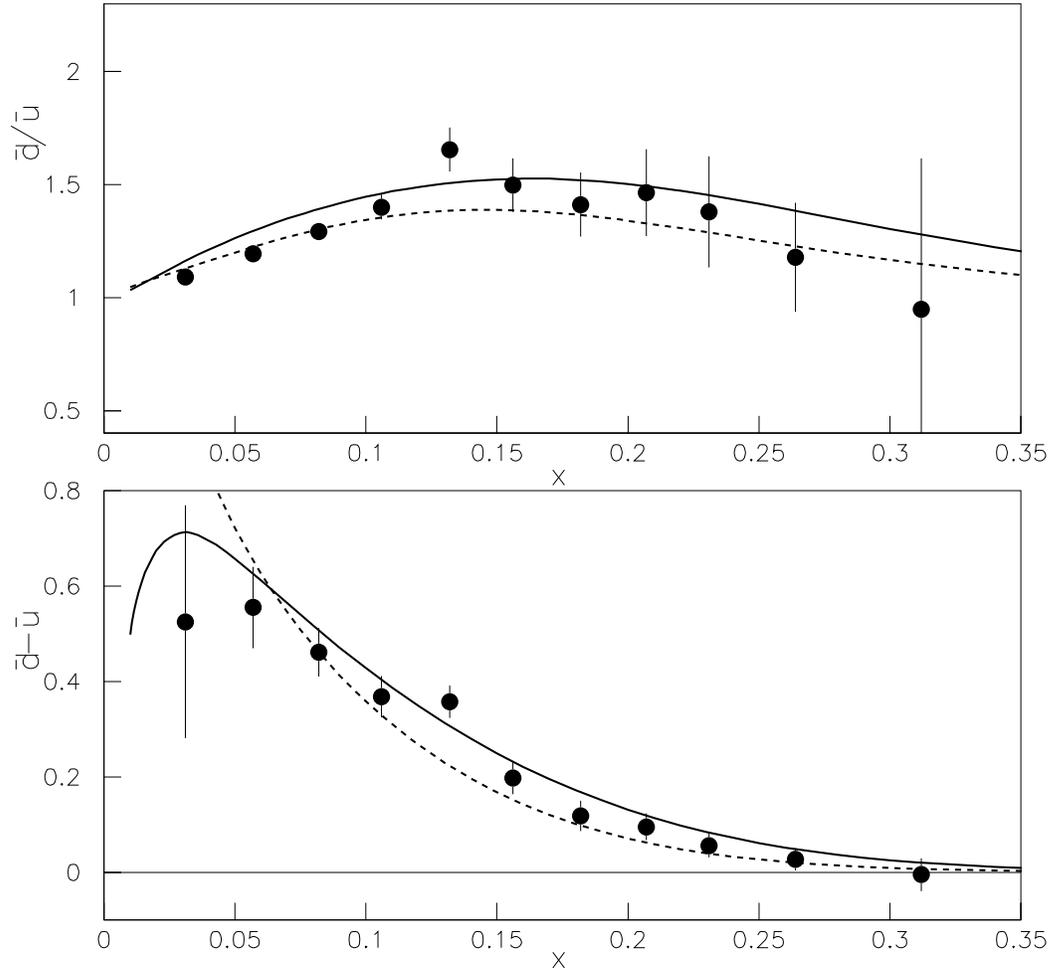,height=6.0in} 
\caption{Same as in Fig.~\ref{fig1} but using a modified valence 
quark distribution in pions with an extra power of $x$ 
(normalized accordingly). See discussion in the text.}
\label{fig2} 
\end{figure}

As can be seen in Fig.~\ref{fig1}, the results of the model 
are impressive, considering that we are representing {\it both} the difference
and the ratio. Nevertheless, in the small-$x$ region the 
model seems to overestimate the value of \ud  due to the steep 
growth of the valence quark distribution of the pion 
as $x\rightarrow 0$. If for instance, we multiply the quark distribution
of the pion by a power of $x$,\footnote{For simplicity we used a power 1,
but other values close to this can do the job as well.} the signaled excessive 
growth at very small $x$ is corrected while the rest of the curve does not 
appreciably changes.
The \ud difference predicted by the model at the valon
scale $Q_v$ thus presents an inflection point about $x \sim 0.05$ and goes to
zero with $x$. The description of the \ud data is thus improved at the
price of having modified the low $x$ behavior of the valence quark
distributions in pions, a region where they are not well known.
In addition, we also get a more accurate description 
of the \udq data in all the measured region (see Fig.~\ref{fig2}). 
It should be noted that similar results are obtained 
by using the low $Q^2$ pion valence quark distributions of
Ref.~\cite{suecos}, calculated with a Monte Carlo based model.
 
It is instructive to look at the integrals of the 
non-perturbative $\bar{u}$ and $\bar{d}$ distributions in order to get an 
idea of the relative weights of the $\left|\pi N\right>$ and 
$\left|\pi \Delta\right>$ fluctuations in the model. 
By fixing the normalization of the bound states to 
fit the experimental data, for the unevolved curves in Fig.~\ref{fig1} 
(Fig.~\ref{fig2}) we have 
$\int_0^1{dx \;\bar{u}^{NP}(x)} \sim 0.28\; (0.15)$ and 
$\int_0^1{dx \;\bar{d}^{NP}(x)} \sim 0.47\; (0.29)$. 
Accordingly, the value of $\int_0^1{dx \; [\bar{d}^{NP}(x) - 
\bar{u}^{NP}(x)]}$ predicted by the model is $0.19$ (0.14)
\footnote{Notice that, as an integral of a non-singlet quantity, 
$\int_0^1{dx \; [\bar{d}(x) - \bar{u}(x)}]$ is independent of 
$Q^2$~\cite{eichten}. Then, our results at the valon scale 
remain unchanged after QCD evolution.}.
This is in good agreement with the experimental result $0.147\pm0.039$, 
as given by the NMC \cite{nmc}.

If, on the other hand, we consider the definition of $\bar{d}(x)-\bar{u}(x)$ 
as given by eq.~(\ref{eq9a}), our prediction of 
$\int_0^1{dx \; [\bar{d}^{NP}(x) - \bar{u}^{NP}(x)]} $ is $0.091~(0.083)$, 
in close agreement with $0.10\pm 0.018$, obtained by the \ns 
Collaboration~\cite{nusea}. Note that this value of the integral is 
significantly lower than the previous one, which we obtained by direct 
integration of the difference between eqs.~(\ref{eq8a}) and (\ref{eq8}). 
This discrepancy is due to the modulation introduced by the CTEQ4M 
$\bar{u}(x) + \bar{d}(x)$ distribution 
used by the \ns Collaboration to extract the \ud  distribution
\footnote{See also Ref.~\cite{nusea} for an additional discussion about 
the discrepancies between \ns and NMC results.}. 
 
A similar analysis of the \ns data has been recently performed 
in the framework of a light cone form factor version of the pion 
cloud model~\cite{mst}. Predictions of this version of the PCM are 
however not very close to the data. 
One reason may be the use of unnatural hard pion distributions in 
$\left|\pi N\right>$ and $\left|\pi \Delta\right>$ fluctuations, which 
produce large contributions to the $\bar{u}$ and $\bar{d}$ 
distributions beyond $x \sim 0.25$. This drawback in the prediction
of \ud translates into the growing behavior of the resulting \udq ratio. 
To obtain an improved description of both \ud and  
\udq within this approach, the addition of an {\em ad-hoc} 
parametrization of the Pauli exclusion principle is needed.  
In particular, in Ref.\cite{mst}, the Pauli effect is normalized to 7$\%$ 
while the total pion cloud contribution to just 5$\%$. It means that
the Pauli contribution would amount to a 58$\%$ of the total asymmetry.
This is a major contrast between this approach and the present work.

Summarizing, we have shown that, including perturbative 
and effective degrees of freedom in a recombination scheme, 
a pion cloud model alone closely describes the recent data of the
\ns Collaboration. With just two parameters, the normalization of the 
$\left|\pi N \right>$ and $\left|\pi \Delta\right>$ fluctuations, 
we have presented an accurate prediction of the flavor asymmetry 
in the light nucleon sea. Remarkably, our model results allow an excellent
fit of both distributions, difference and ratio, in a consistent way.
Finally, we have also signaled a possible 
reason for the apparent discrepancy between \ns and NMC results on the 
GSR violation.

~

Note added: After the conclusion of this paper another PCM evaluation
of both difference and ratio has been performed \cite{alberg}. 
In contrast to ours it is based on the use of form factors. For the chosen 
parameters a reasonable fit of the difference is obtained but the model 
predictions for the ratio \udq  do not fit the experimental data.

\section*{Acknowledgments}
The authors are grateful to Centro Brasileiro de Pesquisas 
F\'{\i}sicas (CBPF) for the warm hospitality extended to them 
during this work. A. Szczurek and F. Steffens are acknowledged for 
useful comments.
J.M. and H.R.C. are supported by Funda\c{c}\~ao de 
Amparo \`a Pesquisa do Estado de Rio de Janeiro (FAPERJ).


\end{document}